\documentclass[12pt,english]{article}
\usepackage[T1]{fontenc}
\usepackage{xcolor}
\usepackage{amsmath}
\usepackage{amssymb}
\usepackage{graphicx}

\makeatletter
\usepackage{jheppub}
\def\@fpheader{\relax}

\allowdisplaybreaks[1]

\makeatother

\usepackage{babel}
\begin{document}
\subheader{}

\title{
Effective GUP-modified Raychaudhuri equation and black hole singularity: four models
}

\author[a]{Keagan Blanchette,}
\author[b]{Saurya Das,}
\author[a]{Saeed Rastgoo}
\affiliation[a]{Department of Physics and Astronomy, York University\\ 4700 Keele Street,Toronto, Ontario M3J 1P3, Canada} 
\affiliation[b]{Department of Physics and Astronomy, University of Lethbridge,\protect\\ 4401 University Drive, Lethbridge, Alberta, Canada, T1K 3M4}
\emailAdd{kblanch@yorku.ca}
\emailAdd{saurya.das@uleth.ca}
\emailAdd{srastgoo@yorku.ca}

\abstract{The classical Raychaudhuri equation predicts the formation
of conjugate points for a congruence of geodesics, in a finite proper
time. This in conjunction with the Hawking-Penrose singularity theorems
predicts the incompleteness of geodesics and thereby the singular
nature of practically all spacetimes. We compute the generic corrections
to the Raychaudhuri equation in the interior of a Schwarzschild black
hole, arising from modifications to the algebra inspired by the generalized
uncertainty principle (GUP) theories. Then we study four specific
models of GUP, compute their effective dynamics as well as their expansion
and its rate of change using the Raychaudhuri equation. We show that
the modification from GUP in two of these models, where such modifications
are dependent of the configuration variables, lead to finite Kretchmann
scalar, expansion and its rate, hence implying the resolution of the
singularity. However, the other two models for which the modifications
depend on the momenta still retain their singularities even in the
effective regime. 

}
\maketitle

\section{Introduction}


As is well-known, most reasonable classical spacetimes are singular,
in the sense of geodesic incompleteness, as predicted by the celebrated
Hawking-Penrose singularity theorems. The essential ingredient behind the formulation of these theorems,
namely the Raychaudhuri equation, predicts the convergence of geodesics
in a finite proper time, and this leads directly to their incompleteness
\cite{Raychaudhuri:1953yv,Penrose:1964wq,Hawking:1969sw}. 

The above singularity being classical however, 
it is expected that it will be resolved by a consistent theory of Quantum Gravity (QG). 
This is particularly true for
black holes and in particular the Schwarzschild
model. While classically a physical singularity exists in the interior
of this black hole, the hope is that quantum gravity effects will lead
to its resolution. This issue has been studied in various approaches
to quantum gravity, in particular, in loop quantum gravity (LQG),
which is a nonperturbative canonical approach to quantization of gravity
\cite{Thiemann:2007pyv}. Within LQG, both the interior and the full
spacetime of Schwarzschild and also lower dimensional black holes
have been studied (see. e.g., \cite{Bojowald:2004af,Ashtekar:2005qt,Bojowald:2005cb,Bohmer:2007wi,Corichi:2015xia,Barrau:2018rts,Alesci:2019pbs,Aruga:2019dwq,Bodendorfer:2019cyv,Brahma:2014gca,Campiglia:2007pb,Chiou:2008nm,Corichi:2015vsa,Cortez:2017alh,Gambini:2008dy,Gambini:2009ie,Gambini:2011mw,Gambini:2013ooa,Gambini:2020nsf,Husain:2004yz,Kelly:2020uwj,Thiemann:1992jj,Campiglia:2007pr,Gambini:2009vp,Rastgoo:2013isa,Corichi:2016nkp,Morales-Tecotl:2018ugi,Blanchette:2020kkk}
and the references within). If one only considers the interior, then
the metric mimics the metric of the Kantowski-Sachs (KS) cosmological
model and one is dealing with a minisuperspace model, meaning a gravitational
system with finite dimensional classical phase space. Within LQG,
this model is usually quantized using polymer quantization \cite{Ashtekar:2002sn,Corichi:2007tf,Morales-Tecotl:2016ijb,Tecotl:2015cya,Flores-Gonzalez:2013zuk}
by first symmetry reducing the model at the classical level and then
applying the quantization procedure (although other works, such as
\cite{Alesci:2019pbs}, exist in which reduction is done after quantization).
The polymer quantization introduces a (set of) parameter(s) into the
theory called the polymer scale. These parameters set minimal scales
of the model which determine the onset of quantum gravitational effects.
These works show a general effective way of avoiding the singularity.

There has been a phenomenological approach to studying certain problems in QG, via the so-called
{\it Generalized Uncertainty Principle} (GUP). Various approaches to QG, black hole physics etc.  
predict the existence of a minimum measurable length and/or a maximum measurable angular momentum. 
For example, examining string theory and its related scattering amplitudes beyond the Planck scale strongly suggests such a length \cite{Amati:1988tn,Gross:1987ar} as does some other approaches to quantum gravity.
This leads to a deformation of the standard Heisenberg commutation relation, which in turn induces correction terms to practically all quantum mechanical Hamiltonians. This leads to  QG effects in a range of systems from the laboratory based to the astrophysical, including potentially measurable ones in the context of 
black holes and cosmology
\cite{AMELINO_CAMELIA_2002,Amelino_Camelia_2001,Cort_s_2005,Pikovski_2012,Marin:2013pga,Bawaj_2015,Amelino_Camelia_2013,Hossenfelder_2013,Amati:1988tn,Kempf:1994su,Maggiore:1993rv,Maggiore_1994,Adler_2001,Das_2008,Nozari_2008,Alonso_Serrano_2018,Myung_2007,Bargue_o_2015,Gangopadhyay_2018,Ali_2014,Ali_2015,Das_2011,Sprenger_2011,Ali_2011,Rovelli:1994ge,Garay:1994en,Scardigli:1999jh,AmelinoCamelia:2008qg,Bosso:2018ckz,Bosso:2017hoq,Bosso:2018uus,Bosso:2019ljf,Casadio_2020,Bosso:2020aqm,Todorinov_2019,Bosso_2020,Bosso_2021,Bonder:2017ckx,Das_2020,das2021bounds,Garcia-Chung:2020zyq,Stargen_2019,das2021test}. 
However, GUP in the context of the Raychaudhuri equation, its deformations and the subsequent implications for singularity resolution, to the best of our knowledge has not been studied extensively. We investigate this further in this article. 
The role of GUP in the interior of black holes has been investigated recently in 
\cite{Bosso:2020ztk,preparation1}. 
Corrections to the Raychaudhuri equation from other sources and its implications to singularity resolution in quantum gravity and cosmology was studied in 
\cite{Das_2014,Das_2015,Burger_2018,Das_2019}.

In this paper, we investigate the modified dynamics of the interior
of the Schwarzschild black hole using Ashtekar-Barbero variables but
using modified algebra inspired by GUP. We consider a generic class
of deformations of the Poisson algebra assuming that such modification
are the phenomenological result of similar modifications at the quantum
level. Using this modified algebra, we derive the dynamics of the
generic equations of motion of the interior and based on that find
the expansion $\theta$ and its rate of change $\frac{d\theta}{d\tau}$
(with $\tau$ being the proper time) using the Raychaudhuri equation.
Then, we discuss the general conditions under which $\theta$ and
$\frac{d\theta}{d\tau}$ remain finite everywhere in the interior.
The finiteness of these quantities implies that no caustic points
for congruence of geodesics, and consequently no singularity, exists.
We then choose four specific subcases of this generic class of models
in which the modifications are either linear or quadratic in configuration
variable or the momenta. We derive the detailed dynamics of each case
as well as the explicit expression for $\theta$ and $\frac{d\theta}{d\tau}$
in relevant cases. We then show that in two of the four cases in which
the modifications depend on the configuration variables, the Kretchmann
scalar, $\theta$ and $\frac{d\theta}{d\tau}$ remain finite everywhere
in the interior, which implies the resolution of the singularity.
However, in the two other cases in which the modifications depend
on the momenta, the Kretchmann scalar diverges even in the effective
regime and the singularity persists. Hence, for the latter two cases
we do not compute $\theta$ and $\frac{d\theta}{d\tau}$.

The structure of this manuscript is as follows. In Sec. \ref{sec:Class-Sch-interior}
we review the dynamics of the interior of the Schwarzschild black
hole in the classical regime using the Ashtekar-Barbero variables.
In Sec. \ref{sec:Class-Ray}, we briefly discuss the Raychaudhuri
equation, its significance and its classical expression and behavior
for the interior of the Schwarzschild black hole. In Sec. \ref{sec:General-GUP},
we introduce the generic class of the GUP modifications we are considering
and derive the generic form of $\theta$ and $\frac{d\theta}{d\tau}$
for this class using the generic dynamics of the interior and the
Raychaudhuri equation. We also discuss the conditions under which
$\theta$ and $\frac{d\theta}{d\tau}$ remain finite. In Sec. \ref{sec:Specific-models}
we consider four specific models within the generic class mentioned.
These are the most common models used in GUP. We analyze both the
dynamics and the behavior of $\theta$ and $\frac{d\theta}{d\tau}$
in these models and show that in two of them the singularity is resolved
while in the other two it persists even in the effective regime. Finally,
in Sec. \ref{sec:Conclusion} we summarize our work and conclude and
also discuss some possible future directions. 

\section{Classical Schwarzschild interior and its dynamics\label{sec:Class-Sch-interior}}

It is well-known that by switching the coordinates $t$ and $r$ in
the metric of the Schwarzschild black hole
\begin{equation}
ds^{2}=-\left(1-\frac{2GM}{r}\right)dt^{2}+\left(1-\frac{2GM}{r}\right)^{-1}dr^{2}+r^{2}d\Omega^{2},
\end{equation}
one can obtain the metric of the interior as
\begin{equation}
ds^{2}=-\left(\frac{2GM}{t}-1\right)^{-1}dt^{2}+\left(\frac{2GM}{t}-1\right)dr^{2}+t^{2}d\Omega^{2},\label{eq:sch-inter}
\end{equation}
where now $t$ is the Schwarzschild interior time that takes values
in the range $t\in(0,2GM)$. In this form, $t$ is the radius of the
2-spheres inside the black hole. The above metric is a special case
of a Kantowski-Sachs (KS) cosmological spacetime that is given by
the metric \cite{Collins:1977fg} 
\begin{equation}
ds_{KS}^{2}=-N(T)^{2}dT^{2}+g_{xx}(T)dx^{2}+g_{\theta\theta}(T)d\theta^{2}+g_{\phi\phi}(T)d\phi^{2}.\label{eq:K-S-gener}
\end{equation}
Note that $x$ here can be a rescaling of the coordinate $r$ in (\ref{eq:sch-inter}),
and $T$ is the generic KS time corresponding to the choice of he
lapse $N(T)$. The KS spacetime is a spatially homogeneous but anisotropic
model. Its spatial hypersurfaces have topology $\mathbb{R}\times\mathbb{S}^{2}$,
and its symmetry group is the KS isometry group $\mathbb{R}\times SO(3)$.

We are interested in expressing the model in terms of the Ashtekar-Barbero
connection $A_{a}^{i}$ and its conjugate, the desitized triad $\tilde{E}_{i}^{a}$.
It turns out that due to the symmetry of the model, $A_{a}^{i},\,\tilde{E}_{i}^{a}$
adapted to this spacetime can be written as \cite{Ashtekar:2005qt}
\begin{align}
A_{a}^{i}\tau_{i}dx^{a}= & \frac{c}{L_{0}}\tau_{3}dx+b\tau_{2}d\theta-b\tau_{1}\sin\theta d\phi+\tau_{3}\cos\theta d\phi,\label{eq:A-AB}\\
\tilde{E}_{i}^{a}\tau_{i}\partial_{a}= & p_{c}\tau_{3}\sin\theta\partial_{x}+\frac{p_{b}}{L_{0}}\tau_{2}\sin\theta\partial_{\theta}-\frac{p_{b}}{L_{0}}\tau_{1}\partial_{\phi},\label{eq:E-AB}
\end{align}
where $b$, $c$, and their respective momenta $p_{b}$ and $p_{c}$,
are functions that only depend on time, and $\tau_{i}=-i\sigma_{i}/2$
are a $su(2)$ basis satisfying $\left[\tau_{i},\tau_{j}\right]=\epsilon_{ij}{}^{k}\tau_{k}$,
with $\sigma_{i}$ being the Pauli matrices. Hence $b,\,c$ comprise
the components of $A_{a}^{i}$ and $p_{b},\,p_{c}$ make up the components
of $\tilde{E}_{i}^{a}$. The parameter $L_{0}$ here is called the
fiducial length. Due to the topology of the model and the presence
of a noncompact direction $x\in\mathbb{R}$ in space, the symplectic
form $\int_{\mathbb{R}\times\mathbb{S}^{2}}\text{d}^{3}x\,\text{d}q\wedge\text{d}p$,
which is necessary to express the Poisson algebra, diverges. Therefore,
in order to cure this one needs to choose a finite fiducial volume
over which the integral is calculated \cite{Ashtekar:2005qt}. This
is a common practice in the study of homogeneous minisuperspace models.
One then introduces an auxiliary length $L_{0}$ to restrict the noncompact
direction to an interval $x\in\mathcal{I}=[0,L_{0}]$. The volume
of the fiducial cylindrical cell in this case becomes $V_{0}=a_{0}L_{0}$,
where $a_{0}$ is the area of the 2-sphere $\mathbb{S}^{2}$ in $\mathcal{I}\times\mathbb{S}^{2}$. 

As usual in gravity, the classical Hamiltonian is the sum of constraints
that generate spacetime diffeomorphisms and internal or Gauss (in
our case $su(2)$) symmetry. The full classical Hamiltonian constraint
in Ashtekar-Barbero formulation is \cite{Thiemann:2007pyv} 
\begin{equation}
H_{\textrm{full}}=\frac{1}{8\pi G}\int d^{3}x\frac{N}{\sqrt{\det|\tilde{E}|}}\left\{ \epsilon_{i}^{jk}F_{ab}^{i}\tilde{E}_{j}^{a}\tilde{E}_{k}^{b}-2\left(1+\gamma^{2}\right)K_{[a}{}^{i}K_{b]}^{j}\tilde{E}_{i}^{a}\tilde{E}_{j}^{b}\right\} ,\label{eq:Full-H-gr-class}
\end{equation}
where $K_{a}^{i}$ is the extrinsic curvature of foliations, $\epsilon_{ijk}$
is the totally antisymmetric Levi-Civita symbol, and $F=dA+A\wedge A$
is the curvature of the Ashtekar-Barbero connection. By replacing
Eqs. (\ref{eq:A-AB}) and (\ref{eq:E-AB}) into (\ref{eq:Full-H-gr-class}),
one obtains the symmetry reduced Hamiltonian of the KS model in $b,\,c,\,p_{b},\,p_{c}$
as \cite{Ashtekar:2005qt,Chiou:2008nm,Corichi:2015xia,Bohmer:2007wi,Morales-Tecotl:2018ugi}
\begin{equation}
H=-\frac{N}{2G\gamma^{2}}\left[2bc\sqrt{p_{c}}+\left(b^{2}+\gamma^{2}\right)\frac{p_{b}}{\sqrt{p_{c}}}\right].\label{eq:H-class-N}
\end{equation}
Given the homogeneous nature of the model, the diffeomorphism constraint
is trivially satisfied, and after imposing the Gauss constraint, one
is left only with the classical Hamiltonian constraint (\ref{eq:H-class-N}).

The classical algebra of the canonical variables also turns out to
be
\begin{equation}
\{c,p_{c}\}=2G\gamma,\quad\quad\{b,p_{b}\}=G\gamma.\label{eq:classic-PBs-bc}
\end{equation}
Considering $q_{ab}$ as the spatial part of the KS metric (\ref{eq:K-S-gener}),
and noticing
\begin{equation}
qq^{ab}=\delta^{ij}\tilde{E}_{i}^{a}\tilde{E}_{j}^{b},
\end{equation}
one can obtain the relations between the KS spatial metric components
and $b,\,c,\,p_{b},\,p_{c}$ as
\begin{align}
g_{xx}\left(T\right)= & \frac{p_{b}\left(T\right)^{2}}{L_{0}^{2}p_{c}\left(T\right)},\label{eq:grrT}\\
g_{\theta\theta}\left(T\right)= & \frac{g_{\phi\phi}\left(T\right)}{\sin^{2}\left(\theta\right)}=p_{c}\left(T\right).\label{eq:gththT}
\end{align}
Note that the lapse $N(T)$ is not determined and can be chosen freely
based on a specific situation. The adapted metric using (\ref{eq:grrT})
and (\ref{eq:gththT}) then becomes 
\begin{equation}
ds^{2}=-N(T)^{2}dT^{2}+\frac{p_{b}^{2}}{L_{0}^{2}\,p_{c}}dx^{2}+p_{c}(d\theta^{2}+\sin^{2}\theta d\phi^{2}).
\end{equation}
Comparing this with the Schwarzschild metric (\ref{eq:sch-inter})
with time $t$ and corresponding lapse, we obtain
\begin{align}
N\left(t\right)= & \left(\frac{2GM}{t}-1\right)^{-\frac{1}{2}},\label{eq:Sch-corresp-1}\\
g_{xx}\left(t\right)= & \frac{p_{b}\left(t\right)^{2}}{L_{0}^{2}p_{c}\left(t\right)}=\left(\frac{2GM}{t}-1\right),\label{eq:Sch-corresp-2}\\
g_{\theta\theta}\left(t\right)= & \frac{g_{\phi\phi}\left(t\right)}{\sin^{2}\left(\theta\right)}=p_{c}\left(t\right)=t^{2}.\label{eq:Sch-corresp-3}
\end{align}
This shows that 
\begin{align}
p_{b}= & 0, & p_{c}= & 4G^{2}M^{2}, &  & \textrm{on the horizon\,}t=2GM,\label{eq:t-horiz}\\
p_{b}\to & 0, & p_{c}\to & 0, &  & \textrm{at the singularity\,}t\to0.\label{eq:t-singular}
\end{align}
In order to understand the physical interpretation of these variables,
we first note from (\ref{eq:Sch-corresp-3}) that $p_{c}$ is the
square of the radius of the infalling 2-spheres. $p_{b}$ is also
related to the areas $A_{x,\theta}$ and $A_{x,\phi}$ of the surfaces
bounded by $\mathcal{I}$ and a great circle along a longitude of
$V_{0}$, and $\mathcal{I}$ and the equator of $V_{0}$, respectively
via \cite{Chiou:2008nm}
\begin{equation}
A_{x,\theta}=A_{x,\phi}=2\pi L_{0}\sqrt{g_{xx}g_{\Omega\Omega}}=2\pi p_{b}.
\end{equation}
In order to better understand the role of $b,\,c$, let us choose
a lapse $N=1$. This is always possible since $N$ is a gauge that
is related to the choice of hypersurface foliations and physics is
invariant under such choice of gauge. The time corresponding to this
lapse is the proper time $\tau$ which has a relation with the generic
time $T$ for the metric (\ref{eq:K-S-gener}),
\begin{equation}
d\tau^{2}=N(T)^{2}dT^{2}.\label{eq:t-tau-N}
\end{equation}
Using the form of the lapse function (\ref{eq:t-tau-N}), we can derive
the equations of motion for $b,\,c$ as \cite{Chiou:2008nm,Blanchette:2020kkk,Bosso:2020ztk}
\begin{align}
b= & \frac{\gamma}{2}\frac{1}{\sqrt{p_{c}}}\frac{dp_{c}}{d\tau}=\gamma\frac{d}{d\tau}\sqrt{g_{\Omega\Omega}}=\frac{\gamma}{\sqrt{\pi}}\frac{d}{d\tau}\sqrt{A_{\theta,\phi}},\\
c= & \gamma\frac{d}{d\tau}\left(\frac{p_{b}}{\sqrt{p_{c}}}\right)=\gamma\frac{d}{d\tau}\left(L_{0}\sqrt{g_{xx}}\right).
\end{align}
These show that, classically, $b$ is proportional to the rate of
change of the square root of the physical area of $\mathbb{S}^{2}$,
and $c$ is proportional to the rate of change of the physical length
of $\mathcal{I}$. 

To obtain the classical dynamics of the interior, we now choose a
different gauge
\begin{equation}
N\left(T\right)=\frac{\gamma\sqrt{p_{c}\left(T\right)}}{b\left(T\right)}.\label{eq:lapsNT}
\end{equation}
The advantage of this lapse function is that the equations of motion
of $c,\,p_{c}$ decouple from those of $b,\,p_{b}$ as we will see
in a moment and it makes it possible to solve them. Using (\ref{eq:lapsNT}),
the Hamiltonian constraint (\ref{eq:H-class-N}) becomes
\begin{equation}
H=-\frac{1}{2G\gamma}\left[\left(b^{2}+\gamma^{2}\right)\frac{p_{b}}{b}+2cp_{c}\right].\label{eq:H-class-1}
\end{equation}
The equations of motion corresponding to this Hamiltonian are
\begin{align}
\frac{db}{dT}= & \left\{ b,H\right\} =-\frac{1}{2}\left(b+\frac{\gamma^{2}}{b}\right),\label{eq:EoM-cls-b}\\
\frac{dp_{b}}{dT}= & \left\{ p_{b},H\right\} =\frac{p_{b}}{2}\left(1-\frac{\gamma^{2}}{b^{2}}\right),\label{eq:EoM-cls-pb}\\
\frac{dc}{dT}= & \left\{ c,H\right\} =-2c,\label{eq:EoM-cls-c}\\
\frac{dp_{c}}{dT}= & \left\{ p_{c},H\right\} =2p_{c}.\label{eq:EoM-cls-pc}
\end{align}
These equations are to be supplemented with the on-shell condition
of the vanishing of the Hamiltonian constraint (\ref{eq:H-class-1})
on the constraint surface\footnote{Here $\approx$ stands for weak equality, i.e., on the constraint
surface.} 
\begin{equation}
\left(b^{2}+\gamma^{2}\right)\frac{p_{b}}{b}+2cp_{c}\approx0.\label{eq:H-weak-vanish}
\end{equation}
Solving these equations one obtains expressions in time $T$. It turns
out that in order to write the solution in Schwarzschild time $t$,
one needs to make the transformation $T=\ln(t)$ in the solutions.
This way one obtains \cite{Ashtekar:2005qt,Bohmer:2007wi,Corichi:2015xia,Blanchette:2020kkk,Bosso:2020ztk}
\begin{align}
b\left(t\right)= & \pm\gamma\sqrt{\frac{2GM}{t}-1},\label{eq:sol-cls-b}\\
p_{b}\left(t\right)= & lL_{0}t\sqrt{\frac{2GM}{t}-1},\label{eq:sol-cls-pb}\\
c\left(t\right)= & \mp\frac{\gamma GMlL_{0}}{t^{2}},\label{eq:sol-cls-c}\\
p_{c}\left(t\right)= & t^{2}.\label{eq:sol-cls-pc}
\end{align}

\begin{figure}
\begin{centering}
\includegraphics[scale=0.7]{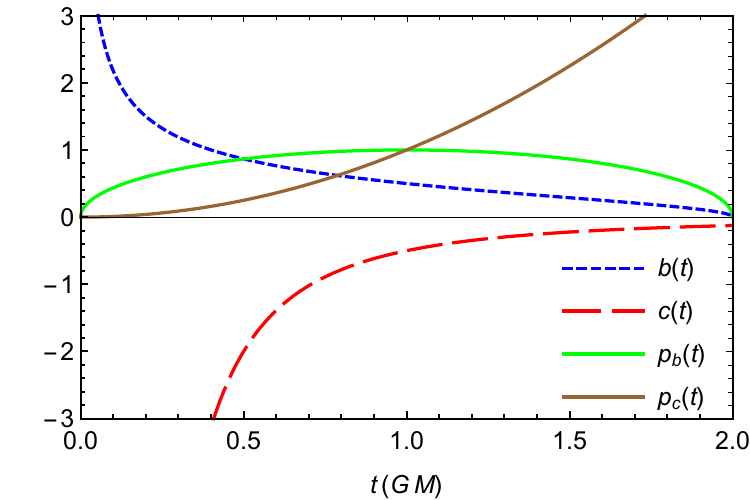}
\par\end{centering}
\caption{The behavior of canonical variables as a function of the Schwarzschild
time $t$. We have chosen the positive sign for $b$ and negative
sign for $c$. The figure is plotted using $\gamma=0.5,\,M=1,\,G=1$
and $L_{0}=1$. \label{fig:class-var-behv}}
\end{figure}

The behavior of these solutions as a function of $t$ is depicted
in Fig. \ref{fig:class-var-behv}. From these equations or the plot,
one can see that $p_{c}\to0$ as $t\to0$, i.e., at the classical
singularity. As a result Riemann invariants such as the Kretschmann
scalar 
\begin{equation}
K=R_{abcd}R^{abcd}\propto\frac{1}{p_{c}^{3}},\label{eq:Kretch}
\end{equation}
all diverge, signaling the presence of a physical singularity for
$p_{c}\to0$ as expected.

\section{The classical Raychaudhuri Equation\label{sec:Class-Ray}}

Let a congruence of (a collection of nearby) geodesics be defined
by the velocity field tangent to the geodesics, $u^{a}(x)$. Then
taking the derivative of $u_{a;b}$ with respect to the proper time
$\tau$ (or affine parameter), we get 
\begin{align}
\frac{du_{a;b}}{d\tau}=u_{a;b;c}\,u^{c}= & \left[u_{a;c;b}+R_{cba}^{d}u_{a}\right]u^{c}\nonumber \\
= & (u_{a;c}u^{c})_{;b}-u_{;b}^{c}u_{a;c}+R_{cbad}u^{c}u^{d}.\label{eq:RE1}
\end{align}
Next, defining the induced metric $h_{ab}=g_{ab}-u_{a}u_{b}$, decomposing
$u_{a;b}$ into its trace, symmetric and antisymmetric parts as follows
$u_{a;b}=\frac{1}{3}\theta h_{ab}+\sigma_{ab}+\omega_{ab}$ and taking
the trace of Eq. (\ref{eq:RE1}), we get 
\begin{equation}
\frac{d\theta}{d\tau}=-\frac{1}{3}\,\theta^{2}-\sigma_{ab}\sigma^{ab}+\omega_{ab}\omega^{ab}-R_{ab}u^{a}u^{b}.\label{eq:RE2}
\end{equation}
Here $\theta$ is the expansion, $\sigma_{ab}\sigma^{ab}$ is the
shear, $\omega_{ab}\omega^{ab}$ is the vorticity term and $R_{ab}$
is the Ricci tensor. As can be seen, most of the terms in the RHS
of the above equation are negative and therefore for a congruence
of geodesics with no vorticity, the above equation can be integrated
to give $\tau<3/\theta_{0}$, where $\theta_{0}$ is the initial value
of $\theta$ and $\tau$ signifies the proper time of geodesic convergence.
In the next few sections we will show how quantum corrections will
introduce positive terms in the RHS of Eq. (\ref{eq:RE2}).

\begin{figure}
\begin{centering}
\includegraphics[scale=0.54]{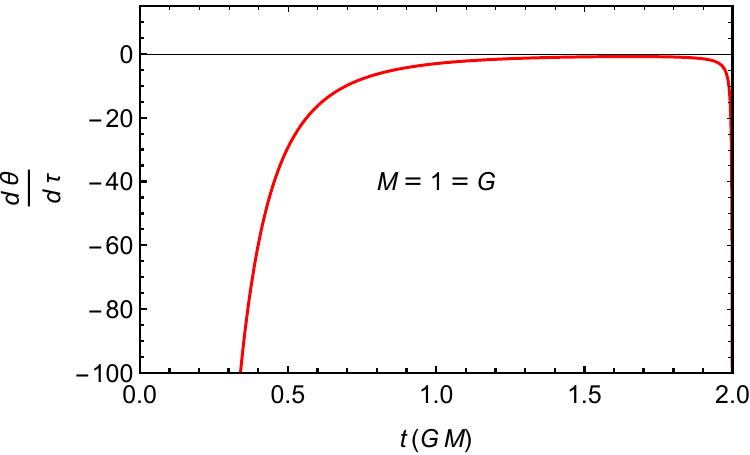} ~~~~~~~~\includegraphics[scale=0.51]{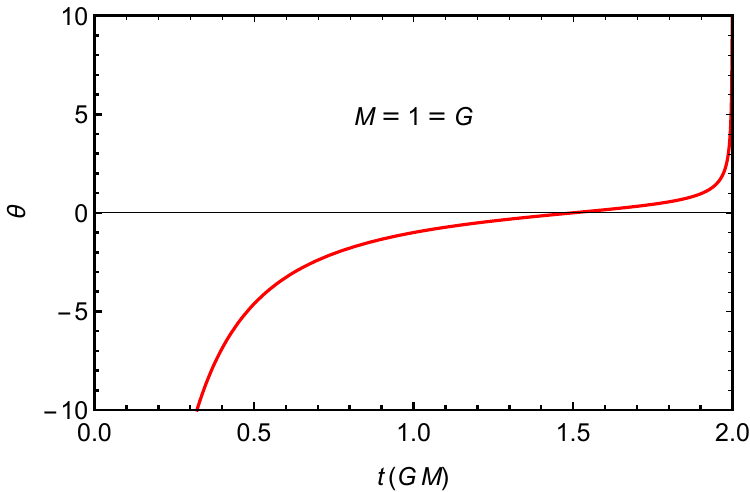}
\par\end{centering}
\caption{Left: $\frac{d\theta}{d\tau}$ as a function of the Schwarzschild
time $t$. Right: negative branch of $\theta$ as a function of $t$.
Both $\theta$ and $\frac{d\theta}{d\tau}$ diverge as we approach
$t\to0$. Note that the divergence at the horizon is due to the choice
of Schwarzschild coordinate system. \label{fig:theta-RE-t-class}}
\end{figure}

Since we consider our model in vacuum, we can set $R_{ab}=0$ in (\ref{eq:RE2}).
Also, in general in KS models, the vorticity term is only nonvanishing
if one considers metric perturbations \cite{Collins:1977fg}. Hence,
$\omega_{ab}\omega^{ab}=0$ in our model, too. This reduces the Raychaudhuri
equation for our analysis to 
\begin{equation}
\frac{d\theta}{d\tau}=-\frac{1}{3}\theta^{2}-\sigma_{ab}\sigma^{ab}.\label{eq:RE}
\end{equation}
It is well-known that the expansion and shear for this model can be written
in terms of $N,\,p_{b},\,p_{c}$ and their time derivatives as \cite{Collins:1977fg,Blanchette:2020kkk,Bosso:2020ztk}
\begin{align}
\theta= & \frac{\dot{p}_{b}}{Np_{b}}+\frac{\dot{p}_{c}}{2Np_{c}},\label{eq:expansion}\\
\sigma^{2}= & \frac{2}{3}\left(-\frac{\dot{p}_{b}}{Np_{b}}+\frac{\dot{p}_{c}}{Np_{c}}\right)^{2}.\label{eq:shear}
\end{align}
Replacing (\ref{eq:EoM-cls-pb}), (\ref{eq:EoM-cls-pc}) and (\ref{eq:lapsNT})
into (\ref{eq:expansion}) and (\ref{eq:shear}) and substituting
them into (\ref{eq:RE}) we obtain \cite{Blanchette:2020kkk,Bosso:2020ztk}
\begin{equation}
\frac{d\theta}{d\tau}=-\frac{1}{2p_{c}}\left(1+\frac{9b^{2}}{2\gamma^{2}}+\frac{\gamma^{2}}{2b^{2}}\right).\label{eq:Class-RE}
\end{equation}
Using (\ref{eq:sol-cls-pc}) and (\ref{eq:sol-cls-b}) in the above
yields \cite{Blanchette:2020kkk,Bosso:2020ztk}
\begin{equation}
\frac{d\theta}{d\tau}=\frac{-2t^{2}+8GMt-9G^{2}M^{2}}{\left(2GM-t\right)t^{3}}.\label{eq:RE-RHS-t-class}
\end{equation}
In the same way, one can obtain 
\begin{equation}
\theta=\pm\frac{1}{2\sqrt{p_{c}}}\left(\frac{3b}{\gamma}-\frac{\gamma}{b}\right)=\pm\frac{-2t+3GM}{t^{\frac{3}{2}}\sqrt{(2GM-t)}}.\label{eq:theta-class}
\end{equation}
These expressions and their plot in Fig. \ref{fig:theta-RE-t-class}
clearly signal the presence of a singularity at $t\to0$ as expected.

\section{General deformed algebra, Effective dynamics and the Raychaudhuri
equation\label{sec:General-GUP}}

As mentioned in the Introduction, 
various approaches to QG, black
hole physics etc. strongly suggest the 
existence of a minimum measurable length
in spacetime. This is often associated with the Planck length, but in principle can be any length scale lying between the Planck and the electroweak scale. 
This gives rise to an 
effective and generic 
modification of the standard Heisenberg algebra. 
Inspired by the above, and the fact that 
a corrected quantum algebra also implies suitable modifications of the corresponding 
 Poisson algebra, we propose the following fundamental Poisson brackets between the canonical variables as

\begin{align}
\{b,p_{b}\} & =G\gamma F_{1}\left(b,c,p_{b},p_{c},\beta_{b},\beta_{c}\right),\\
\{c,p_{c}\} & =2G\gamma F_{2}\left(b,c,p_{b},p_{c},\beta_{b},\beta_{c}\right),
\end{align}
where the modifications are encoded entirely in $F_{1}$ and $F_{2}$,
and hence the non-deformed classical limit is obtained by setting
$F_{1}=1=F_{2}$. Such modification will result in the effective equations
of motion 
\begin{align}
\frac{db}{dT} & =\{b,H\}=-\frac{1}{2}\left(b+\frac{\gamma^{2}}{b}\right)F_{1},\label{eq:EoM-gen-b}\\
\frac{dp_{b}}{dT} & =\{p_{b},H\}=\frac{p_{b}}{2}\left(1-\frac{\gamma^{2}}{b^{2}}\right)F_{1},\label{eq:EoM-gen-pb}\\
\frac{dc}{dT} & =\{c,H\}=-2cF_{2},\label{eq:EoM-gen-c}\\
\frac{dp_{c}}{dT} & =\{p_{c},H\}=2p_{c}F_{2}.\label{eq:EoM-gen-pc}
\end{align}
As before, these equations should be supplemented by weakly vanishing
of the Hamiltonian constraint (\ref{eq:H-class-1}). 

From the above equations of motion for $b,\,p_{b}$, we can infer
\begin{equation}
\frac{db}{dp_{b}}=\frac{\left(\gamma^{2}+b^{2}\right)}{\left(\gamma^{2}-b^{2}\right)}\frac{b}{p_{b}}.\label{eq:db-dpb}
\end{equation}
which leads to 
\begin{equation}
p_{b}=\frac{Ab}{\gamma^{2}+b^{2}},\label{eq:pb-in-b}
\end{equation}
with $A$ being a constant of integration. In the same way from the
equations of motion for $c,\,p_{c}$, we get 
\begin{equation}
\frac{dc}{dp_{c}}=-\frac{c}{p_{c}},\label{eq:dc-dpc}
\end{equation}
which yields 
\begin{equation}
p_{c}=\frac{B}{c},\label{eq:pc-in-c}
\end{equation}
with $B$ being another integration constant. From the last two equations
we can also deduce a couple of basic results that will be useful later.
First, note that if one demands that the Kretchmann scalar (\ref{eq:Kretch})
remains finite everywhere inside the black hole, then $p_{c}$ should
remain finite everywhere, and particularly at $t\to0$. Hence, from
Eq. (\ref{eq:pc-in-c}) and assuming a finite $p_{c}$ everywhere
in the interior, we deduce that $c$ should remain finite everywhere
in the interior too. Second, from Eq. (\ref{eq:pb-in-b}) we can have
three types of behaviors for $b(t)$, particularly at $t\to0$, as
follows: 
\begin{enumerate}
\item If for $t\to0$ we get $b\to0$, then $p_{b}\to 0$ too in that
region. 
\item If $b$ remains finite, then $p_{b}$ will remain finite. 
\item If $b\to\pm\infty$, then $p_{b}\to0$. 
\end{enumerate}
The above equations of motion (\ref{eq:EoM-gen-b})-(\ref{eq:EoM-gen-pc})
can now be substituted into the Raychaudhuri equation, Eq. (\ref{eq:RE})
and (\ref{eq:expansion}) to obtain (with $N=\frac{\gamma\sqrt{p_{c}}}{b}$
as before): 
\begin{equation}
\frac{d\theta}{d\tau}=\frac{1}{4\gamma^{2}p_{c}}\left(2\gamma^{2}F_{1}^{2}+4b^{2}F_{1}F_{2}-4\gamma^{2}F_{1}F_{2}-b^{2}F_{1}^{2}-12b^{2}F_{2}^{2}-\frac{F_{1}^{2}\gamma^{4}}{b^{2}}\right),\label{eq:REfg}
\end{equation}
and 
\begin{equation}
\theta=\pm\frac{1}{2\gamma\sqrt{p_{c}}}\left(bF_{1}-\frac{\gamma^{2}F_{1}}{b}+2bF_{2}\right),\label{eq:theta-factored}
\end{equation}
in terms of the canonical variables. We need both $\theta$ and $\frac{d\theta}{d\tau}$
to remain finite everywhere, particularly close to and at the singularity.
Since we are assuming $p_{c}|_{t\to0}\to\mathrm{finite}$ due to requirement
for finiteness of the Kretchmann scalar at the singularity, only the
terms inside the parentheses in $\theta$ and $\frac{d\theta}{d\tau}$
above matter.

In what follows, we will consider four cases of linear modifications
to the Poisson algebra. These cases, as suggested by literature in
the field, are the most used cases in GUP-inspired models. These cases
include the configuration-dependent modifications 
\begin{align}
F_{1}(q,p)= & 1+\alpha_{b}b, & F_{2}(q,p)= & 1+\alpha_{c}c,\label{eq:config-dep-1}\\
F_{1}(q,p)= & 1+\beta_{b}b^{2}, & F_{2}(q,p)= & 1+\beta_{c}c^{2},
\end{align}
and the momentum-dependent modifications 
\begin{align}
F_{1}(q,p)= & 1+\alpha_{b}^{\prime}p_{b}, & F_{2}(q,p)= & 1+\alpha_{c}^{\prime}p_{c},\\
F_{1}(q,p)= & 1+\beta_{b}^{\prime}p_{b}^{2}, & F_{2}(q,p)= & 1+\beta_{c}^{\prime}p_{c}^{2}.
\end{align}
In what follows we consider the effect of such modifications on the
dynamics of the interior and the behavior of $\theta$ and $\frac{d\theta}{d\tau}$
in this region.

\section{Specific models\label{sec:Specific-models}}

We consider four distinct GUP inspired models in this section and examine the consequences. 
These four models are chosen because they cover most of the spectrum of GUPs that authors have used to study Planck scale/QG corrections in quantum systems, 
suitably adapted to the problem at hand. Following the 
lead of those works studying linear and quadratic GUP models, our four cases cover the linear and quadratic in the canonical variables $b,c,p_ b$ and $p_c$.

\subsection{Model 1: $F_{1}=1+\beta_{b}b^{2},\,F_{2}=1+\beta_{c}c^{2}$}

This is the case whose dynamics was studies in \cite{Bosso:2020ztk}.
Here, the algebra becomes

\begin{align}
\left\{ b,p_{b}\right\} = & G\gamma\left(1+\beta_{b}b^{2}\right),\label{eqn:b-pb-m1}\\
\left\{ c,p_{c}\right\} = & 2G\gamma\left(1+\beta_{c}c^{2}\right),\label{eqn:c-pc-m1}
\end{align}
and the corresponding equations of motion are 
\begin{align}
\frac{db}{dT}= & \left\{ b,H\right\} =-\frac{1}{2}\left(b+\frac{\gamma^{2}}{b}\right)\left(1+\beta_{b}b^{2}\right),\label{eq:EoM-diff-b-m1}\\
\frac{dp_{b}}{dT}= & \left\{ p_{b},H\right\} =\frac{p_{b}}{2}\left(1-\frac{\gamma^{2}}{b^{2}}\right)\left(1+\beta_{b}b^{2}\right),\label{eq:EoM-diff-pb-m1}\\
\frac{dc}{dT}= & \left\{ c,H\right\} =-2c\left(1+\beta_{c}c^{2}\right),\label{eq:EoM-diff-c-m1}\\
\frac{dp_{c}}{dT}= & \left\{ b,H\right\} =2p_{c}\left(1+\beta_{c}c^{2}\right).\label{eq:EoM-diff-pc-m1}
\end{align}
Once again, these equations should be supplemented by the weakly vanishing
($\approx0$) of the Hamiltonian constraint (\ref{eq:H-class-1}),
\begin{equation}
\left(b^{2}+\gamma^{2}\right)\frac{p_{b}}{b}+2cp_{c}\approx0.\label{eq:weak-van}
\end{equation}
The solutions to these equations of motion in terms of the Schwarzschild
time $t$ are \cite{Bosso:2020ztk}
\begin{align}
b\left(t\right)= & \pm\frac{\gamma\sqrt{2GMt^{\beta_{b}\gamma^{2}}-t\left(2\gamma^{2}GM\right)^{\beta_{b}\gamma^{2}}}}{\sqrt{t\left(2\gamma^{2}GM\right)^{\beta_{b}\gamma^{2}}-2\beta_{b}\gamma^{2}GMt^{\beta_{b}\gamma^{2}}}},\label{eq:b-eff-t-no-L0-m1}\\
p_{b}\left(t\right)= & \frac{\ell_{c}}{\sqrt{-\beta_{c}}}t^{-\beta_{b}\gamma^{2}}\sqrt{\left[2GMt^{\beta_{b}\gamma^{2}}-t\left(2\gamma^{2}GM\right)^{\beta_{b}\gamma^{2}}\right]\left[t\left(2\gamma^{2}GM\right)^{\beta_{b}\gamma^{2}}-2\beta_{b}\gamma^{2}GMt^{\beta_{b}\gamma^{2}}\right]},\label{eq:pb-eff-t-no-L0-m1}\\
c\left(t\right)= & \mp\frac{\ell_{c}}{\sqrt{-\beta_{c}}}\frac{\gamma GM}{\sqrt{t^{4}-\ell_{c}^{2}\gamma^{2}G^{2}M^{2}}},\label{eq:c-eff-t-no-L0-m1}\\
p_{c}\left(t\right)= & \sqrt{t^{4}+\ell_{c}^{2}\gamma^{2}G^{2}M^{2}}.\label{eq:pc-eff-t-no-L0-m1}
\end{align}
where we have set $l=1$. Following \cite{Bosso:2020ztk}, in these
equations we have defined a physical scale 
\begin{equation}
\ell_{c}^{2}=-\beta_{c}L_{0}^{2}.\label{eq:prescript}
\end{equation}
The introduction of this scale is necessary to avoid the dependence
of physical quantities such as expansion and shear on the fiducial
parameter $L_{0}$. Note that if we identify this $p_{c}^{\mathrm{min}}$
with the one derived from LQG in \cite{Corichi:2015xia}, we will
obtain $\ell_{c}^{(\alpha)2}=\Delta$ \cite{Bosso:2020ztk}, where
$\Delta$ is the minimum area in LQG. 

\begin{figure}
\begin{centering}
\includegraphics[scale=0.55]{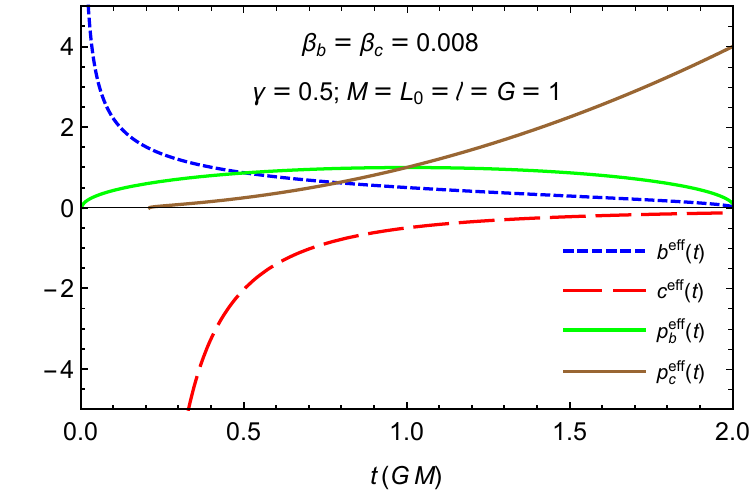}~~~~~~~~\includegraphics[scale=0.55]{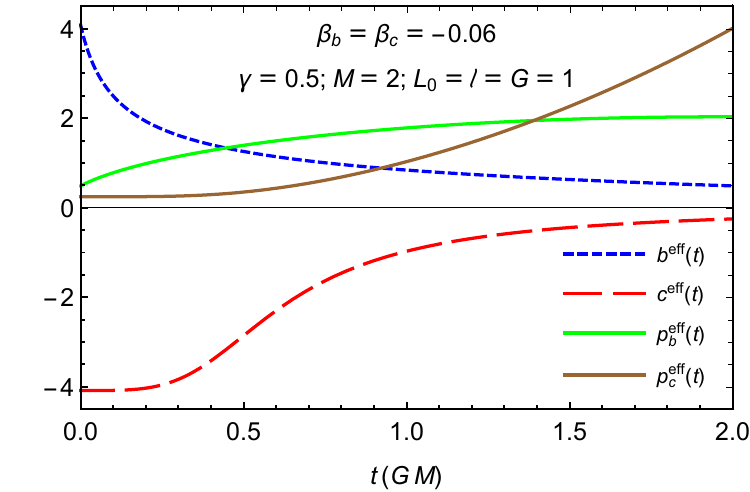}
\par\end{centering}
\begin{centering}
\vspace{10pt}
 \includegraphics[scale=0.55]{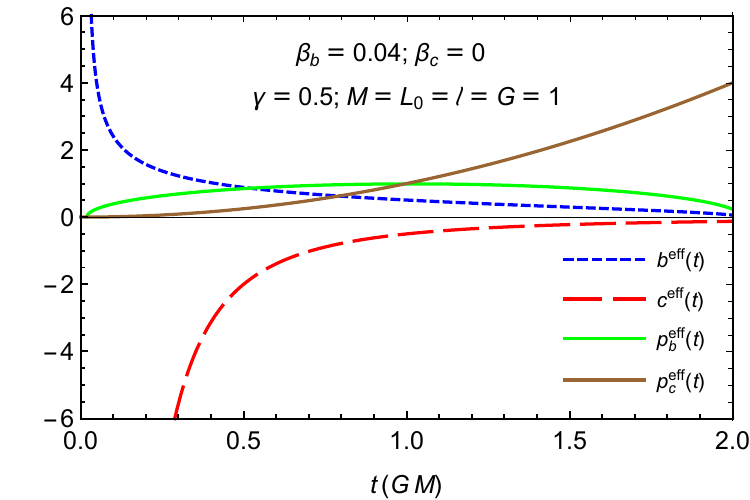}~~~~~~~~\includegraphics[scale=0.55]{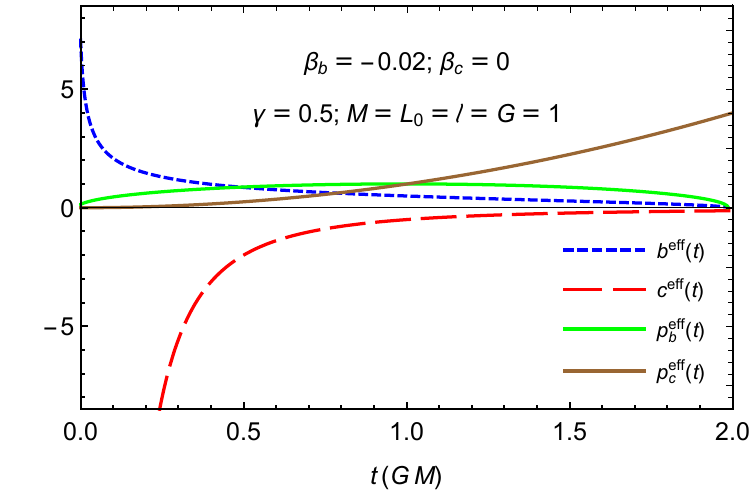}
\par\end{centering}
\begin{centering}
\vspace{10pt}
 \includegraphics[scale=0.55]{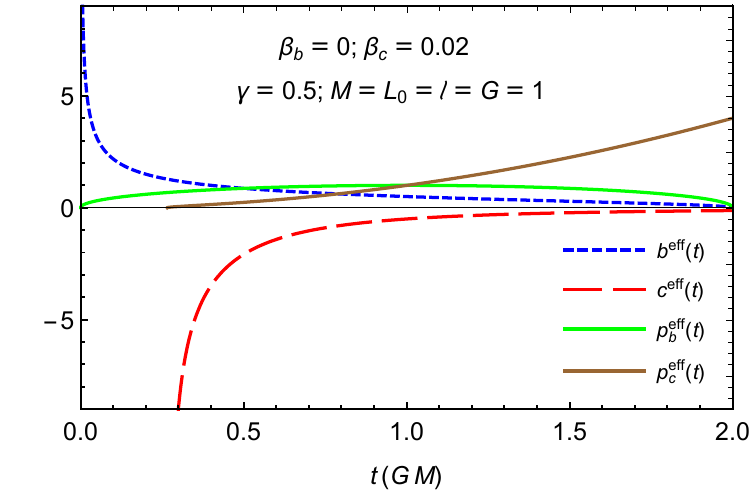}~~~~~~~~\includegraphics[scale=0.55]{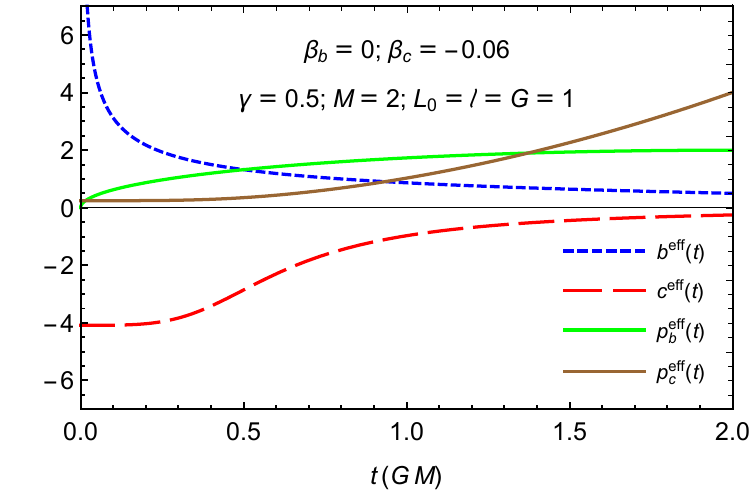}
\par\end{centering}
\caption{The behavior of solutions of the modified case in the Schwarzschild
time $t$ for positive, negative and vanishing $\beta_{b}$ and $\beta_{c}$
for the whole interior. We have chosen the positive sign for $b$
and negative sign for $c$. Note that for nonvanishing negative $\beta_{c}$
we always get a minimum nonvanishing value for $p_{c}$, while a nonvanishing
negative $\beta_{b}$ leads to a finite value of $b$ at $t\to0$.
The values of parameters are mentioned on each plot. \label{fig:GUP-1-pos-neg}}
\end{figure}

The above solutions are plotted in Fig. \ref{fig:GUP-1-pos-neg}.
In general: 
\begin{itemize}
\item If $\beta_{c}<0$ then $p_{c}$ never vanishes, and hence the Kretschmann
scalar does not diverge. Consequently the singularity is resolved
effectively. Also in this case $c$ becomes bounded everywhere in
the interior.
\item If $\beta_{b}<0$ then $b$ is bounded everywhere in the interior.
\item If $\beta_{c}=0$, then $p_{c}\to0$ for $t\to0$ and the Kretchamnn
scalar blows up in that region. Hence, singularity will be still present.
Also in this case $c$ will not be bounded.
\item If $\beta_{b}\geq0$, then $b$ will not be bounded.
\item If $\beta_{c}>0$, then the evolution stops at some point before reaching
$t=0$ due to $p_{c}$ becoming complex. Also (\ref{eq:prescript})
will not make sence for a real scale $\ell_{c}$.
\end{itemize}

Therefore we can conclude that the case of interest for us is the one in
which both $\beta_{b},\,\beta_{c}<0$ (top right plot). In this case
not only $p_{c}$ acquires a minimum value and the Kretchmann scalar
remains finite, but also $b$ and $c$ are bounded.

\begin{figure}
\begin{centering}
\includegraphics[scale=0.55]{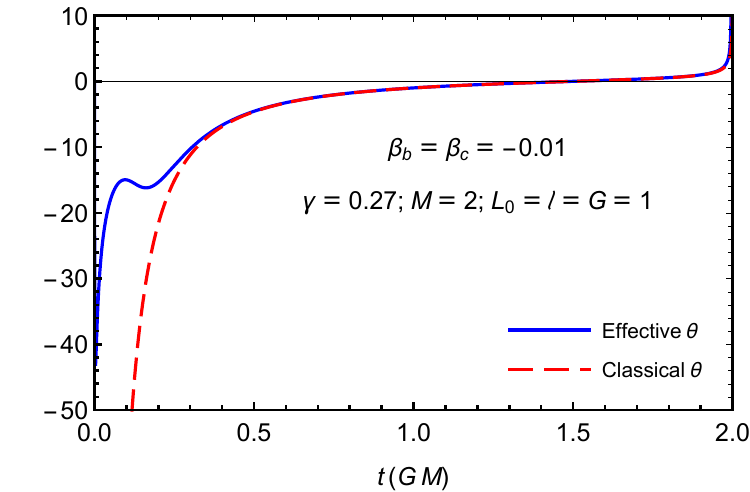}~~~~~~~~
\includegraphics[scale=0.55]{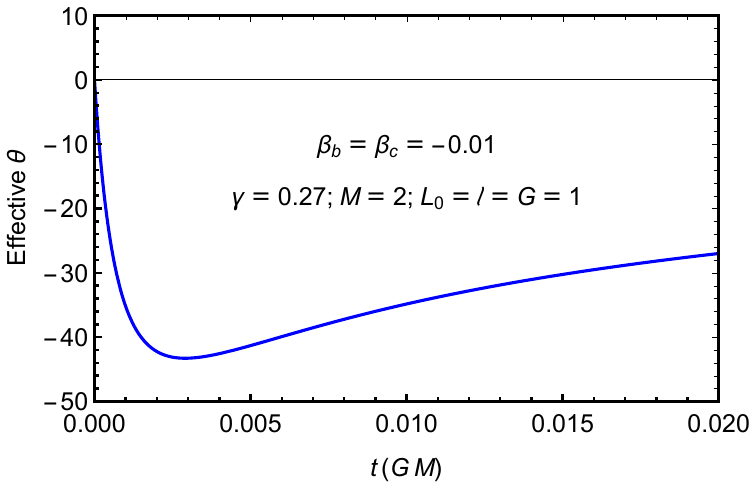}
\par\end{centering}
\begin{centering}
\vspace{10pt}
 \includegraphics[scale=0.55]{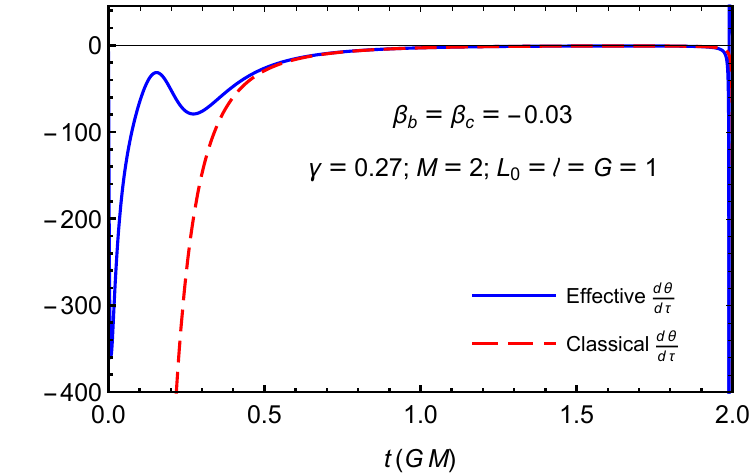}~~~~~~~~\includegraphics[scale=0.55]{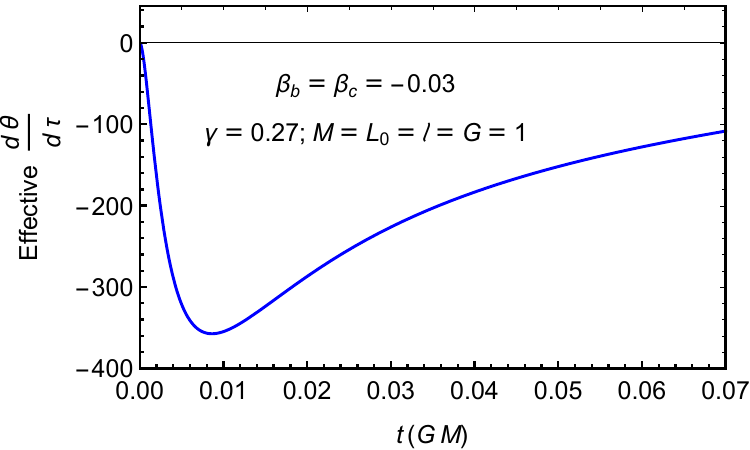}
\\
\par\end{centering}
\caption{Plots of expansion and its rate of change for model 1. Top left: classical
vs effective $\theta$ as a function of the Schwarzschild time $t$.
Top right: closeup of the effective $\theta$ as a function of $t$.
Bottom left: classical vs effective $\frac{d\theta}{d\tau}$ as a
function of $t$. Top right: closeup of the effective $\frac{d\theta}{d\tau}$
as a function of $t$. \label{fig:theta-RE-model-1-t}}
\end{figure}

From the solution (\ref{eq:b-eff-t-no-L0-m1}) (also seen in Fig \ref{fig:GUP-1-pos-neg}),
and assuming since $\beta_{b},\,\beta_{c}<0$, we see that
\begin{align}
b|_{t\to0^{+}}\to & \frac{1}{\sqrt{-\beta_{b}}},\\
F_{1}|_{t\to0^{+}}\to & 0,\\
F_{2}|_{t\to0^{+}}\to & 0.
\end{align}
Considering these limits and looking at (\ref{eq:REfg}) and (\ref{eq:theta-factored}),
we see that both $\theta$ and $\frac{d\theta}{d\tau}$ vanish at
$t\to0$. This in fact can be seen by computing the expression for
the expansion 
\begin{equation}
\theta=\frac{1}{2\gamma\sqrt{p_{c}}}\left[3b-\frac{\gamma^{2}}{b}+\beta_{b}b\left(b^{2}-\gamma^{2}\right)+2\beta_{c}c^{2}b\right],
\end{equation}
and the Raychaudhuri equation \cite{Bosso:2020ztk},
\begin{align}
\frac{d\theta}{d\tau}= & -\frac{9b^{2}}{4\gamma^{2}p_{c}}-\frac{\gamma^{2}}{4b^{2}p_{c}}-\frac{1}{2p_{c}}\nonumber \\
 & +\frac{\beta_{b}}{2\gamma^{2}p_{c}}\left(b^{4}-\gamma^{4}\right)-\frac{\beta_{c}c^{2}}{\gamma^{2}p_{c}}\left(5b^{2}+\gamma^{2}\right)\nonumber \\
 & -\frac{\beta_{b}^{2}b^{2}}{4\gamma^{2}p_{c}}\left(b^{2}-\gamma^{2}\right)^{2}-\frac{3b^{2}\beta_{c}^{2}c^{4}}{\gamma^{2}p_{c}}+\frac{\beta_{b}\beta_{c}b^{2}c^{2}}{\gamma^{2}p_{c}}\left(b^{2}-\gamma^{2}\right),\label{eq:RE-model-1}
\end{align}
for this model, and then replacing in them the solutions (\ref{eq:b-eff-t-no-L0-m1})-(\ref{eq:pc-eff-t-no-L0-m1})
for $\beta_{b},\beta_{c}<0$ and plotting them versus the Schwarzschild
time $t$. These plots are presented in Fig. \ref{fig:theta-RE-model-1-t},
in which one can compare the behavior of effective $\theta$ and $\frac{d\theta}{d\tau}$
versus their classical counterparts. We see that far from the the
position where used to be the classical singularity, the effective
behavior follows the classical one almost identically. However, close
to the $t=0$ region, the defocusing effective corrections dominate
and prevent $\theta$ and $\frac{d\theta}{d\tau}$ from diverging.
This shows that the singularity is resolved in the effective regime.
Furthermore, interestingly $\frac{d\theta}{d\tau}$ shows a similar
qualitative behavior (double bump) as the $\bar{\mu}$ case in (most
of) the loop quantum gravity approach(es) to this model (Fig. 8 in
\cite{Blanchette:2020kkk}).

\subsection{Model 2: $F_{1}=1+\beta_{b}^{\prime}p_{b}^{2},\,F_{2}=1+\beta_{c}^{\prime}p_{c}^{2}$}

In this case, once $F_{1},\,F_{2}$ are replaced into (\ref{eq:EoM-gen-b})-(\ref{eq:EoM-gen-pc}),
it is possible to analytically solve the equation in $c,\,p_{c}$
in Schwarzschild time $t$, 
\begin{align}
c= & GMlL_{0}\gamma\frac{\sqrt{1-\beta_{c}^{\prime}t^{4}}}{t^{2}},\\
p_{c}= & \frac{t^{2}}{\sqrt{1-\beta_{c}^{\prime}t^{4}}},\label{eq:pc-model2}
\end{align}
where first we have solved the differential equations in $T$, replaced
$T\to\ln\left(t\right)$ and then matched the classical limits the
known classical solutions Eqs. (\ref{eq:sol-cls-b})-(\ref{eq:sol-cls-pc}).
Immediately, we see from (\ref{eq:pc-model2}) that 
\begin{equation}
\lim_{t\to0^{+}}p_{c}=0,
\end{equation}
and hence the Kretchmann scalar diverges at $t\to0$ and singularity
is not resolved even in the effective regime. Furthermore $c$ blows
up at $t\to0$. So we will not further analyze the behavior of $\theta$
and $\frac{d\theta}{d\tau}$ in this case.

\subsection{Model 3 $F_{1}=1+\alpha_{b}b,\,F_{2}=1+\alpha_{c}c$ }

For this model, too, it is possible to analytically solve for $c,\,p_{c}$,
while $b,\,p_{b}$ can be obtained numerically. For $c,\,p_{c}$ we
obtain 
\begin{align}
c= & -\frac{GM\gamma lL_{0}}{t^{2}+\alpha_{c}GM\gamma lL_{0}},\\
p_{c}= & t^{2}+\alpha_{c}GM\gamma lL_{0}.
\end{align}
This shows that $p_{c}$ at $t\to0$ acquires a minimum which depends
on $L_{0}$. Once again we can use the prescription introduced in
\cite{Bosso:2020ztk} to define a new physical scale 
\begin{equation}
\ell_{c}^{(\alpha)}=\alpha_{c}L_{0},
\end{equation}
and thus the minimum values of $p_{c}$ becomes 
\begin{equation}
p_{c}^{\mathrm{min}}=\ell_{c}^{(\alpha)}GM\gamma.
\end{equation}
Again, if we identify this $p_{c}^{\mathrm{min}}$ with the one derived
from LQG in \cite{Corichi:2015xia}, we will once again obtain $\ell_{c}^{(\alpha)2}=\Delta$.

Using the solutions above, we see that at $t\to0$ 
\begin{equation}
c=-\frac{1}{\alpha_{c}},
\end{equation}
and hence 
\begin{equation}
F_{2}|_{t\to0}=0.
\end{equation}
Replacing these forms of $F_{1}$ and $F_{2}$ in (\ref{eq:theta-factored})
and (\ref{eq:REfg}) yields 
\begin{align}
\theta|_{t\to0}= & \frac{1}{2\gamma\sqrt{p_{c}}}\left(b^{2}-\gamma^{2}\right)\frac{F_{1}}{b}\nonumber \\
= & \frac{1}{2\gamma\sqrt{p_{c}}}\left(b^{2}-\gamma^{2}\right)\left(\alpha_{b}+\frac{1}{b}\right),
\end{align}
and 
\begin{align}
\frac{d\theta}{d\tau}\bigg|_{t\to0}= & -\frac{1}{4\gamma^{2}p_{c}}\left(b^{2}-\gamma^{2}\right)^{2}\left(\frac{F_{1}}{b}\right)^{2}\nonumber \\
= & -\frac{1}{4\gamma^{2}p_{c}}\left[\left(b^{2}-\gamma^{2}\right)\left(\alpha_{b}+\frac{1}{b}\right)\right]^{2}\nonumber \\
= & -\left[\theta|_{t\to0}\right]^{2}.
\end{align}
It is clear from above two equations that the only way to keep both
$\theta$ and $\frac{d\theta}{d\tau}$ finite is for $b$ to remain
finite at $t\to0$.

We can see these results in another way. By replacing $c$ from (\ref{eq:pc-in-c})
in $F_{2}$ we obtain 
\begin{equation}
F_{2}=1+\alpha_{c}\frac{B}{p_{c}}.
\end{equation}
Substituting both of the above $F_{1},\,F_{2}$ in (\ref{eq:theta-factored})
and (\ref{eq:REfg}) one obtains 
\begin{equation}
\theta=\frac{1}{2\gamma\sqrt{p_{c}}}\left[3b-\frac{\gamma^{2}}{b}+\alpha_{b}\left(b^{2}-\gamma^{2}\right)+\alpha_{c}\frac{2bB}{p_{c}}\right],\label{eq:theta-model-3}
\end{equation}
and 
\begin{align}
\frac{d\theta}{d\tau}= & -\frac{9b^{2}}{4\gamma^{2}p_{c}}-\frac{\gamma^{2}}{4b^{2}p_{c}}-\frac{1}{2p_{c}}\nonumber \\
 & +\frac{\alpha_{b}}{\gamma^{2}bp_{c}}\left(b^{4}-\gamma^{4}\right)-\frac{B\alpha_{c}}{\gamma^{2}p_{c}^{2}}\left(5b^{2}+\gamma^{2}\right)\nonumber \\
 & -\frac{\alpha_{b}^{2}}{4\gamma^{2}p_{c}}\left(b^{2}-\gamma^{2}\right)^{2}-\frac{3b^{2}B^{2}\alpha_{c}^{2}}{\gamma^{2}p_{c}^{3}}+\frac{B\alpha_{b}\alpha_{c}b}{\gamma^{2}p_{c}^{2}}\left(b^{2}-\gamma^{2}\right).\label{eq:RE-model-3}
\end{align}
From these two expressions we see that a necessary and sufficient
condition for finiteness of both $\theta$ and $\frac{d\theta}{d\tau}$
is the finiteness of $b$ when $t\to0$. Also note the similarity
of this expression with Eq. (\ref{eq:RE-model-1}) from model 1. From
the discussion in Sec. \ref{sec:General-GUP}, the finiteness of $b$
means that all the four canonical variables should remain finite at
$t\to0$ for both $\theta$ and $\frac{d\theta}{d\tau}$ not to diverge.
This is in fact the case as we will see below.

\begin{figure}
\begin{centering}
\includegraphics[scale=0.6]{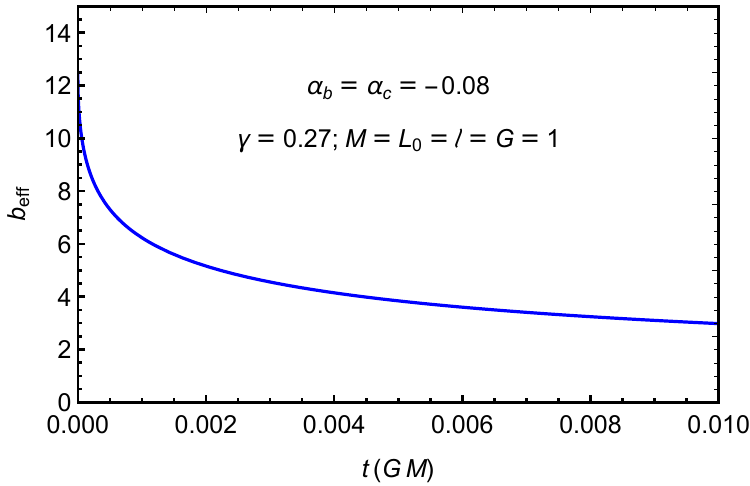}
\par\end{centering}
\caption{Plot of the solution for $b$ as a function of the Schwarzschild time
$t$ for model 3 close to the region that used to be the singularity.
It is clear that $b$ remains finite as $t\to0^{+}$. \label{fig:b-model-3}}
\end{figure}

The above analysis is confirmed by numerical solutions of the differential
equations for $b,\,p_{b}$ in this case. From these numerical solutions,
particularly the one for $b$ which is plotted in Fig \ref{fig:b-model-3},
it is clear that $b$ is bounded in the interior and especially for
$t\to0^{+}$. 

\begin{figure}
\begin{centering}
\includegraphics[scale=0.55]{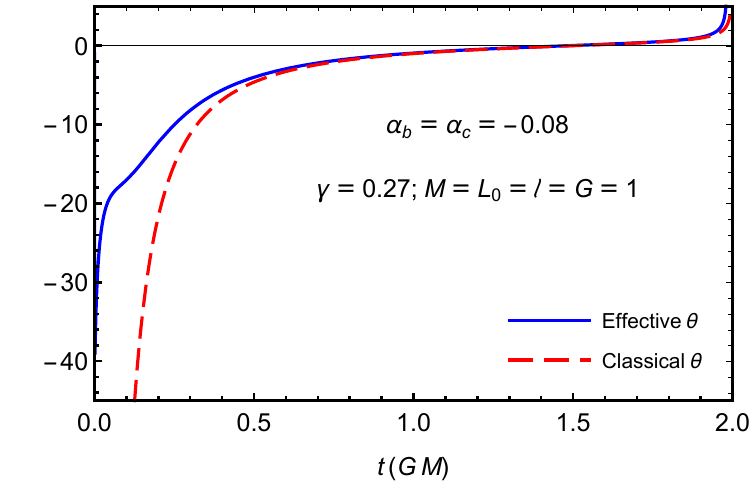}~~~~~~~~
\includegraphics[scale=0.55]{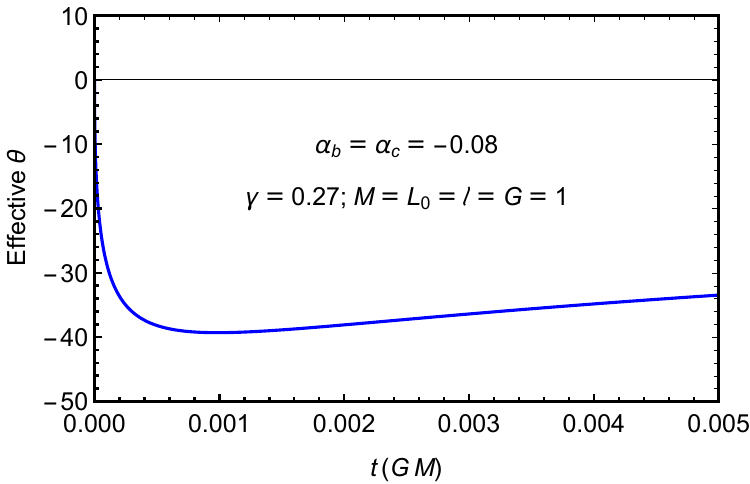}
\par\end{centering}
\begin{centering}
\vspace{10pt}
 \includegraphics[scale=0.55]{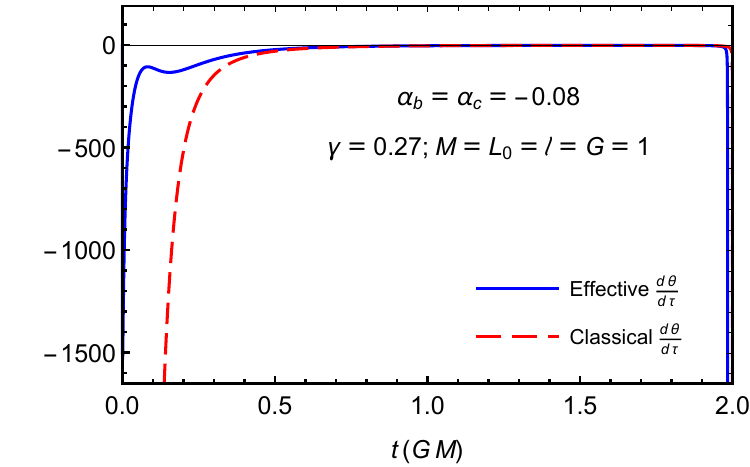}~~~~~~~~\includegraphics[scale=0.55]{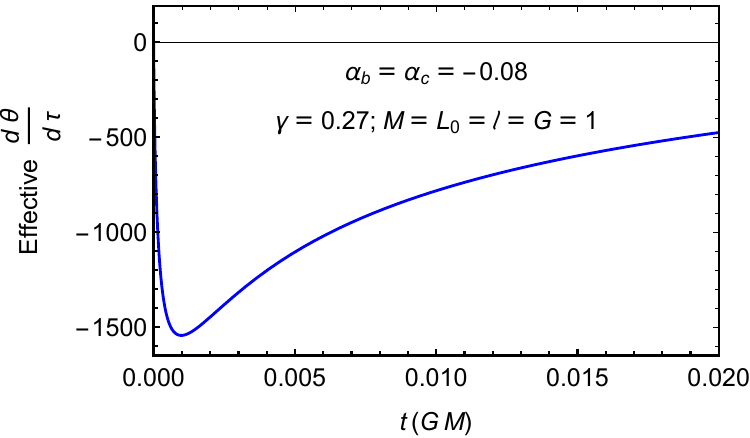}
\\
\par\end{centering}
\caption{Plots of expansion and its rate of change for model 3. Top left: classical
vs effective $\theta$ as a function of the Schwarzschild time $t$.
Top right: closeup of the effective $\theta$ as a function of $t$.
Bottom left: classical vs effective $\frac{d\theta}{d\tau}$ as a
function of $t$. Top right: closeup of the effective $\frac{d\theta}{d\tau}$
as a function of $t$. \label{fig:theta-RE-model-3-t}}
\end{figure}

Furthermore, by using the numerical solutions for $b,\,p_{b}$ and
the analytical solutions for $c,\,p_{c}$ in expressions (\ref{eq:theta-model-3})
and (\ref{eq:RE-model-3}) for $\theta$ and $\frac{d\theta}{d\tau}$,
one can obtain the plot of these quantities. These are presented in
Fig. \ref{fig:theta-RE-model-3-t}. Once again we see that far from
the position where used to be the classical singularity, the classical
and the effective quantities matc almost exactly. However, as $t\to0$,
the effective terms take over and turn the expansion and its rate
toward the value zero. Also note that once again the double bump pattern
is visible in the plot of $\frac{d\theta}{d\tau}$.

\subsection{Model 4: $F_{1}=1+\alpha_{b}^{\prime}p_{b},\,F_{2}=1+\alpha_{c}^{\prime}p_{c}$}

For this case, the solutions to $c,\,p_{c}$ in $t$ are 
\begin{align}
c= & -GM\gamma lL_{0}\frac{1-\alpha_{c}^{\prime}t^{2}}{t^{2}},\\
p_{c}= & \frac{t^{2}}{1-\alpha_{c}^{\prime}t^{2}}.
\end{align}
Hence, similar to Model 2, we have $\lim_{t\to0^{+}}p_{c}=0$ and
the Kretschmann scalar blows up at $t\to0$. Therefore, the singularity
persists even in the effective GUP regime.

\section{Discussion and conclusion\label{sec:Conclusion}}

In this work, we have studied the effects of modifying the Poisson
bracket inspired by GUP on the Raychaudhuri equation in the interior
of the Schwarzschild black hole. This modification leads to an effective
algebra that can be interpreted as a modification inherited from the
quantum algebra. As a result, the equations of motion will be modified
and give us an effective dynamics in the interior of the black hole. 

We have first studied a generic class of modifications and analyzed
the conditions under which the expansion scalar $\theta$ and its
rate of change $\frac{d\theta}{d\tau}$, i.e., the Raychaudhuri equation,
remain finite everywhere inside the black hole. This finiteness signals
the absence of caustic points and particularly in this case, a physical
singularity.

Armed with such a generic analysis, we studied four specific models
that is usually considered in GUP theories with linear or quadratic
mortification to the algebra. We studied their effective dynamics
and analyzed in detail, the behavior of $\theta$ and $\frac{d\theta}{d\tau}$
in each model. We have shown that in two of these models, in which
the modifications are momentum dependent, the singularity persists.
However, in the other two model which are either linearly or quadratically
dependent of the configuration variables, due to quantum gravity correction,
the Kretchamann scalar, $\theta$ and $\frac{d\theta}{d\tau}$ remain
finite everywhere inside the black hole. This is a strong indication
that the singularity of the black hole is resolved effectively. In
addition to being finite, both $\theta$ and $\frac{d\theta}{d\tau}$
approach zero as the Schwarzschild time $t\to0^{}$.

The main reason for the aforementioned behavior of $\theta$ and $\frac{d\theta}{d\tau}$
is the following: in the interior of the Schwarzschild black hole which
is a special form of the Kantowski-Sachs cosmological model, both
$\theta$ and $\frac{d\theta}{d\tau}$ depend on the time derivatives
of the momenta of the model. Replacing these time derivatives from
the classical equations of motion into the expressions for $\theta$
and $\frac{d\theta}{d\tau}$ leads to terms that all have negative
terms, implying focusing of the geodesics which ultimately lead to
caustic points with $\theta,\,\frac{d\theta}{d\tau}\to-\infty$ for
$t\to0^{+}$. This is not surprising given the attractive nature of
gravity. However, repeating the same procedure but now suing the effective
equations of motion leads to two sets of terms. The classical ones
that are all negative as before and terms coming from the modifications
that are positive. These terms are quite small far from $t\to0^{+}$
but dominate and take over close to that region and turn the values
of $\theta$ and $\frac{d\theta}{d\tau}$ over to zero rather than
$-\infty$. One can effective interpret these terms as repulsive. 

Although the models we consider here do not exhaust all possibilities and 
other GUP models can be considered in principle, as mentioned earlier, the models we study here are the ones that are more frequently studied in the literature. Having said that, for the sake of completeness and to shed more light on other GUP models, it is worth examining them in the future.

As a future project, we would like to extend our results to more general
spacetimes. In particular, we would like to study the form of modifications
to a generic class of metrics needed for the singularity to be resolved
especially in relation to the behavior of $\theta$ and $\frac{d\theta}{d\tau}$.

\acknowledgments{

S. D. and S. R. acknowledge the support of the Natural Sciences and
Engineering Research Council of Canada (NSERC), {[}S. D.: funding
reference number RGPIN-2019-05404, S. R. funding reference numbers
RGPIN-2021-03644 and DGECR-2021-00302{]} 

}

\bibliographystyle{JHEP}
\bibliography{mainbib}

\end{document}